\documentclass[3p,times]{elsarticle}
\usepackage{ecrc}
\usepackage{booktabs}
\usepackage{graphicx}
\usepackage{subfigure}
\usepackage{indentfirst}
\usepackage{float}
\usepackage{siunitx}

\volume{00}

\firstpage{1}

\journalname{Physica A}

\runauth{Jun Qin et al.}

\jid{procs}

\jnltitlelogo{Physica A}

\CopyrightLine{2011}{Published by Elsevier Ltd.}

\usepackage{amssymb}

\usepackage[figuresright]{rotating}

\begin{document}

\begin{frontmatter}

\dochead{}

\title{Some Results on Ethnic Conflicts Based on Evolutionary Game Simulation}

\author{Jun Qin$^{1}$, Yunfei Yi$^{4,5}$, Hongrun Wu$^5$, Yuhang Liu$^1$, Xiaonian Tong$^1$, Bojin Zheng*$^{1,2,3}$}

\address{$^1$ College of Computer Science,
South-Central University for Nationalities, Wuhan 430074, China}
\address{$^2$ State Key Laboratory of Networking and Switching Technology, Beijing University of Posts and Telecommunications, Beijing 100876, China}
\address{$^3$ School of Software, Tsinghua University, Beijing 100084, China}
\address{$^4$ College of Computer and Information Science, Hechi University, Yizhou 546300, China}
\address{$^5$ State Key Laboratory of Software Engineering, Wuhan University, Wuhan 430074, China}
\ead{zhengbojin@gmail.com}

\begin{abstract}
The force of the ethnic separatism, essentially origining from negative effect of ethnic identity, is damaging the stability and harmony of multiethnic countries. In order to eliminate the foundation of the ethnic separatism and set up a harmonious ethnic relationship, some scholars have proposed a viewpoint: ethnic harmony could be promoted by popularizing civic identity. However, this viewpoint is discussed only from a philosophical prospective and still lack supports of scientific evidences. Because ethic group and ethnic identity are products of evolution and ethnic identity is the parochialism strategy under the perspective of game theory, this paper proposes an evolutionary game simulation model to study the relationship between civic identity and ethnic conflict based on evolutionary game theory.
The simulation results indicate that: 1) the ratio of individuals with civic identity has a positive association with the frequency of ethnic conflicts; 2) ethnic conflict will not die out by killing all ethnic members once for all, and it also cannot be reduced by a forcible pressure, i.e., increasing the ratio of individuals with civic identity;
3) the average frequencies of conflicts can stay in a low level by promoting civic identity periodically and persistently.
\end{abstract}

\begin{keyword}


Civic identity\sep Ethnic identity\sep Ethnic conflict\sep Evolutionary game simulation
\end{keyword}

\end{frontmatter}


\section{Introduction}

Ethnic groups, a named social group of people based on what is perceived as shared ancestry, cultural traditions, communicational interaction, and history, are the primary entities to provide human's basic needs\cite{weber1978economy}. However, the ethnic group identity, i.e. ethnic identity(dividing the world into categories of `us' and `them'\cite{banks2003ethnicity}), may result in long-term and destructive conflicts between different groups\cite{banks2003ethnicity,Kelman1999-581}.
Kelman et al. pointed out that moral exclusion of ethnic identity, in the extreme, may result in ethnic cleansing and genocide\cite{Kelman2004,kellow1998role}. Rupesinghe and Burton concurred that most ethnic conflicts involved inconsistence of identity\cite{Rupesinghe1987-527,McGarry2013-politics}.
Further, the ethnic identity may evolve into ethnic separatism in multi-ethnic countries, which may lead to ethnic conflicts between ethnic groups or even divide a country into new sovereign territories, such as the Yugoslav wars and Sudan's Darfur conflict etc..
China- a country with 56 ethnic groups, has been threated by the force of ethnic separatism and ethnic conflicts for many years\cite{EYuan2011,YapengWang2002}. Chinese scholars have conducted many related works on the solution of ethnic conflicts and the elimination of negative effect of ethnic identity\cite{LiLiu2011,YiLiu2007,RenhouYang1996}. Most of those scholars concured a viewpoint: ethnic harmony could be promoted by popularizing civic identity. Because civic identity leads people to tolerate others despite of other people's ethnicity \cite{peoples2010humanity,youniss1997we}.
However, this viewpoint was only discussed based on philosophy, and it required sufficient evidences to convince people.

Current papers conducted lots of eminent and significant works on explaining social phenomenon by evolutionary theory. For examples, P. James and D. Goetze explained ethnic conflict and ethnic formation by evolutionary theory\cite{James2001-Evolutionary, goetze1998evolution}; Bester and Guth explained altruistic preferences by evolutionary game theory\cite{Bester1998-193}; Rajiv Sethi and E. Somanathan explained reciprocity as an aspect of preference interdependence by evolutionary game theory\cite{Sethi2001-273}; Choi and Samuel studied the relation of parochial altruism and war by evolutionary simulation\cite{Choi2007-636}.

Inspired by these remarkable works\cite{James2001-Evolutionary, goetze1998evolution,Bester1998-193,Sethi2001-273,Choi2007-636}, in this paper we try to test the viewpoint `ethnic harmony could be promoted by popularizing civic identity' and explore a possible way for the solution of conflicts by a scientific method- based on evolutionary game simulation.
To meet this goal, we firstly map the related concepts(such as ethnic group, ethnic identity, civic identity behaviors etc.) to corresponding concepts in evolutionary game theory. Then we establish a model to simulate the relationship between ethnic conflicts and civic identity. We carried out this model for 30 times.
The simulation results indicate that ethnic conflicts will not die out by cleaning all the ethnic identity members once and for all, though the number of individuals with civic identity has a negative association with the frequencies of conflicts.
Moreover, the frequencies of conflicts cannot be reduced just by increasing the number of individuals with civic identity, because the frequencies will bound back quickly or even higher. However, it is possible to control the ethnic conflicts within a low level by promoting the ratio of civic identity individuals persistently/periodly.

\section{Method}

Based on the current studies on ethnology, evolutionary game theory, and Choi and Samuel's model, we illustrated the relationship of ethnic identity, civic identity and ethnic conflicts by evolutionary game theory.

In this section, firstly we give the foundations to research ethnic conflicts with evolutionary game theory(see section 2.1). Secondly, we elaborate how to map concepts in ethnology to concepts in evolutionary game theory(see section 2.2). Thirdly we show how to establish a evolutionary game model to simulate the relation of civic identity and ethnic conflicts(see section 2.3).
\subsection{The Foundations to study ethnic conflict by evolutionary game theory}

Ethnic groups are shaped by environment, mobility and cultural transmission etc.. Further, ethnic identity comes from cumulative changes of the environment or the evolved mental capacities over years\cite{goetze1998evolution}. That's to say, both ethnic group and ethnic identity are products of evolution\cite{James2001-Evolutionary}. Owing to the evolution, people with the same ethnic identity tend to cooperate intensely with ingroups and to be hostile toward outgroups\cite{James2001-Evolutionary}.
On the contrast, members with a sense of civic identity will risk their lives to protect national interests, and help people without considering their ethnic group, race, or class etc\cite{schwartz2011handbook}.
As the outcome of evolution is to promote individual and group interests(i.e. improving inclusive fitness), some individuals may sacrifice their own well-being and reproductive potential to fight for resources and interests of their group.

All of these behaviors are the game strategies(parochialism and tolerationism strategy etc.) from the perspective of game theory. According to these strategies, groups may cooperate or evoke conflicts with other groups, and both cases are the results of game between ethnic groups.
From this perspective, we can simulate ethnic conflicts with evolutionary game theory.


\subsection{Mapping ethnology to evolutionary game theory}
To simulate ethnic conflicts by evolutionary game theory, we map the related ethnological concepts to corresponding concepts in evolutionary game theory. The mapping is presented as Fig.1.
On the level of inter-group, we map ethnic and civic identity to parochialism and tolerance strategy respectively.
On the level of intra-group, we map the spirit of self-sacrifice and selfish to altruism and non-altruism strategy respectively.
So each individual has four kinds of strategies: parochial altruism, parochial nonaltruism, tolerant altruism, tolerant nonaltruism.

\begin{figure}[H]
\centering\includegraphics[width=10cm,height=5cm]{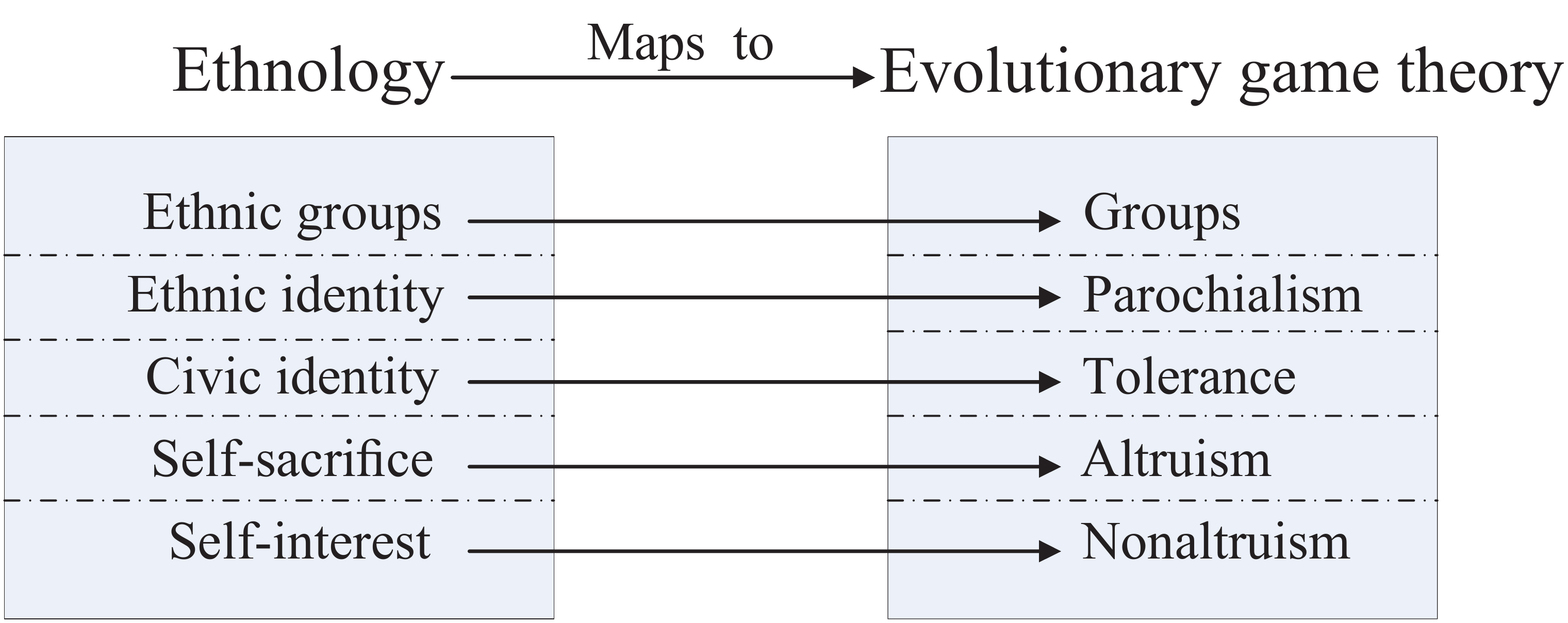}
\caption{ The mapping from concepts in ethnology to concepts in evolutionary game theory}
\label{Fig.1}
\end{figure}

\subsection{Establishing the evolutionary game simulation model}

The same as the previous studies, we use the GA(Genetic Algorithm) to simulate the evolutionary process.
We represent the four types of strategies as two alleles. Therefore, there are four genotypes: parochial altruists(abbr.PA), tolerant altruism (abbr.TA), parochial nonaltruism(abbr.PN), and tolerant nonaltruism (abbr.TN).
Moreover, individuals with genotype P(PA or PN)- abbreviated as P individuals, are hostile toward other groups, while T individuals(TA or TN) are friendly to outgroups.
Similarly, A individuals(PA or TA) contribute interests to the public within one group, while N individuals(PN or TN) do not contribute interests but grab from A individuals in the group.
When a conflict occurs between two groups, only PA individuals engage in conflict, because the N individuals are not willing to sacrifice to benefit their fellow group members. Yet, when two groups are in peace status, T individuals benefit from the cooperated group according to the payoff table.

Furthermore, our evolutionary game model includes four parts: initialization, inter-group interaction, intra-group interaction and update. The model is shown as Fig.2.
\begin{figure}
\centering\includegraphics[width=8cm,height=10cm]{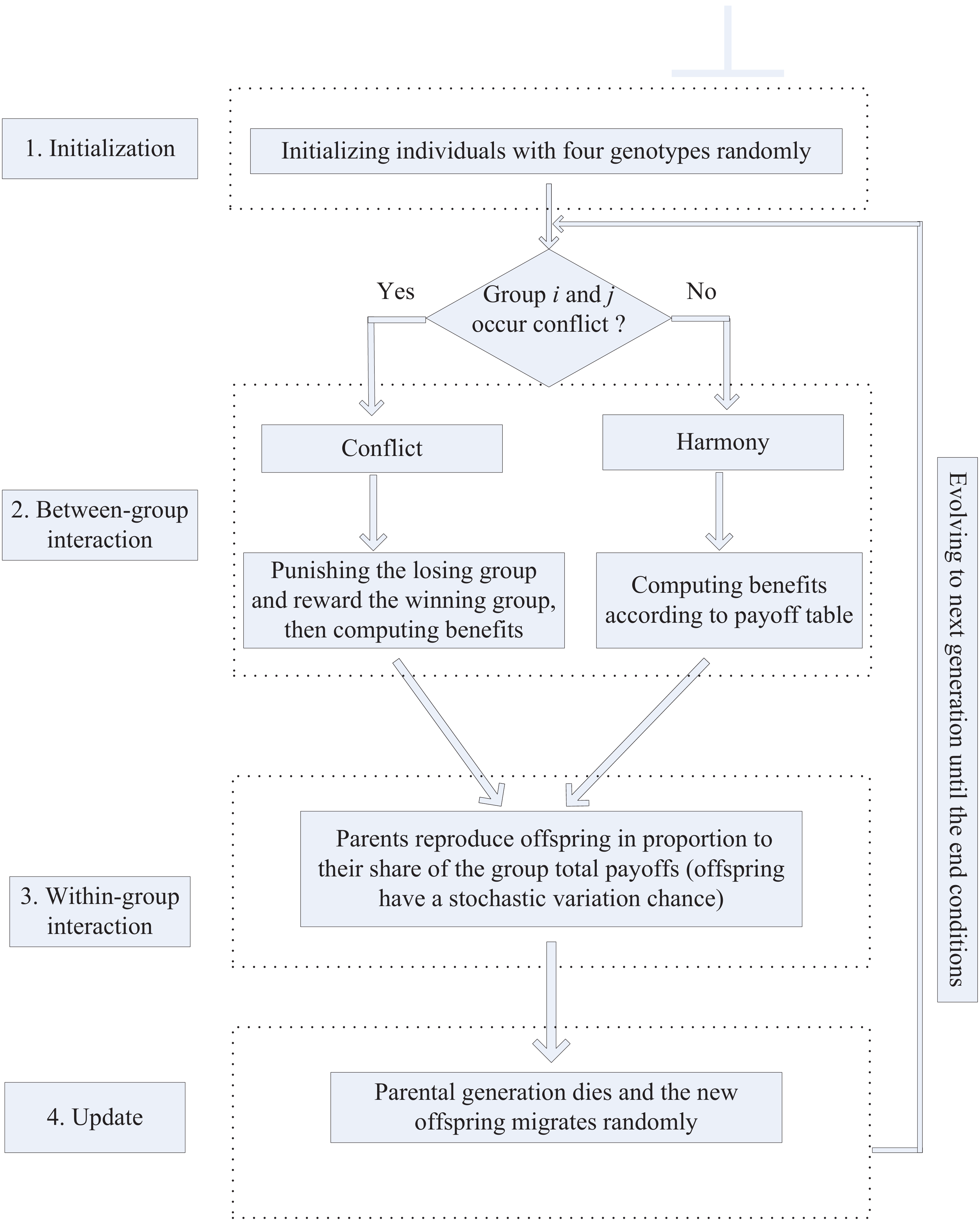}
\caption{The evolutionary game model}
\label{Fig.2}
\end{figure}

\subsubsection{The initialization}

To illustrate the association between civic identity and the frequency of ethnic conflict, the ratio of T individuals and ethnic conflicts need to be calculated in the simulations. Initialization with too many(or too few) T individuals or uneven distribution of genotypes may lead to the statistical error in the simulations. To avoid this initial statistical error, we initialize each group individuals $m$ with four genotypes randomly.

\subsubsection{The inter-group interaction}
In the step of inter-group interaction, a conflict occurs between two arbitrary groups $i$ and $j$ with probability $|p_{ij}|$ (where $p_{ij}=f_i^{PA}-f_j^{PA}$ is the difference between $f_i^{PA}$ and $f_j^{PA}$, which are parochial altruists in group $i$ and $j$ respectively). PA individuals are fighters for their group. So the more PA individuals one group has, the greater probability this group wins a conflict, and the smaller chance the opponent group wins.
If $p_{ij}>0$(i.e. group $i$ has more PA fighters than group $j$), group $i$ has a great probability to win a conflict.
Here, we set group $i$ winning a conflict with probability $0.5+0.5p_{ij}$.
Meanwhile, group $i$ will not win the conflict with a probability $0.5-0.5p_{ij}$, i.e., lose or tie. We set the losing probability and the tie probability equivalently, i.e, both are $0.25-0.25p_{ij}$.
These values of probability are vice verse for $p_{ij}<0$. If $p_{ij}=0$(i.e. the equal number of PA individuals in two groups), the conflict ends with a tie. The probability distribution when $p_{ij}>0$ is listed in table 1.

\begin{figure}[H]
\centering\includegraphics[width=10cm,height=8cm]{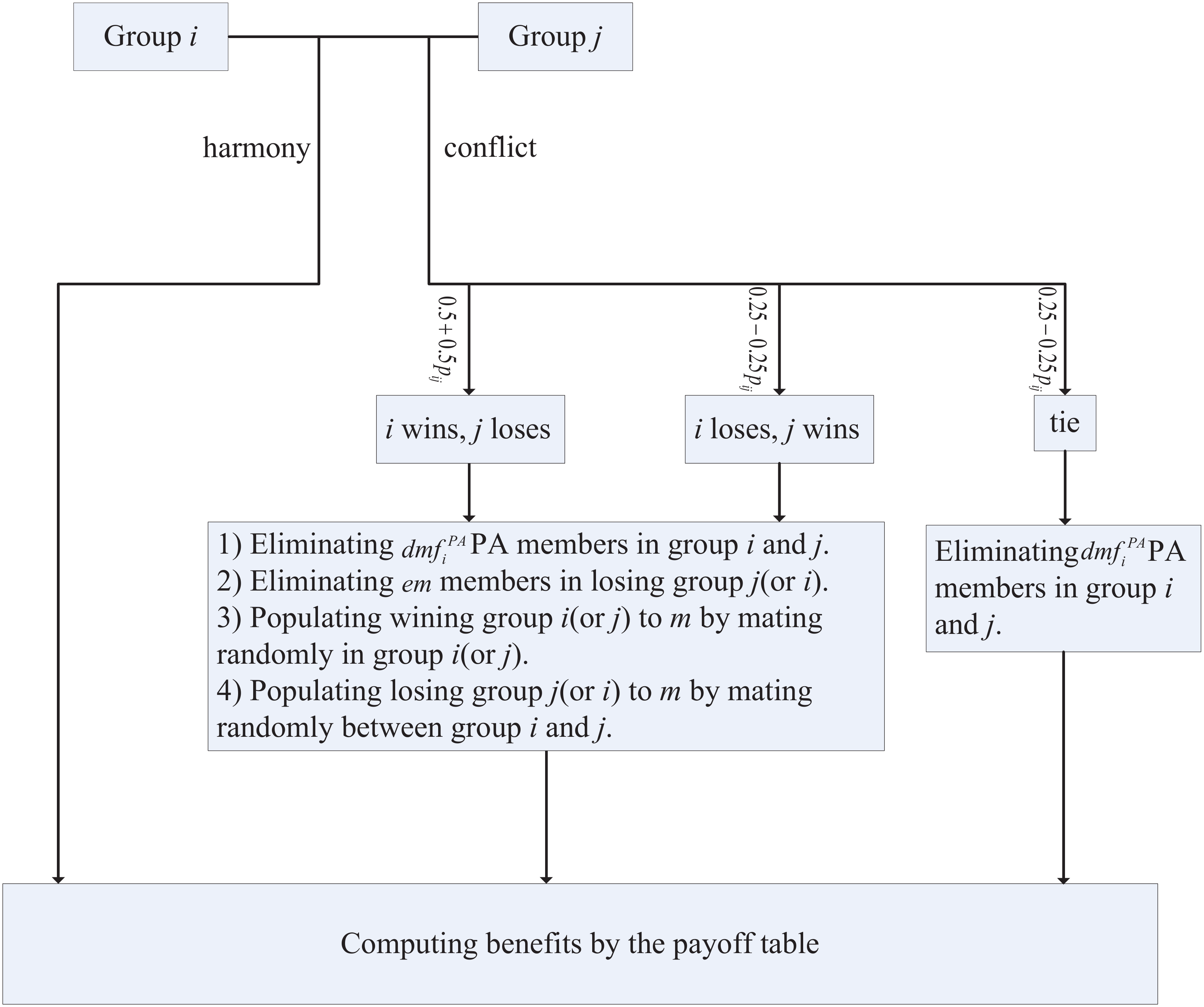}
\caption{Inter-group interaction between group $i$ and $j$ when $p_{ij}>0$}
\label{Fig.btinter}
\end{figure}

If group $i$ wins in the conflict, a ratio $d$ of fighters(PA individuals) in both groups $i$ and $j$ die, and a ratio $e$ of the surviving fighters and nonfighters of the losing group $j$ are killed. Yet only fighters in both groups are eliminated when the conflict ends in a draw. The eliminated individuals in both groups are populated by offspring produced by randomly selected parents in the winning group. Then the two reconstructed groups' benefits are calculated respectively by the payoff table(table 2). However, if the two groups are in a harmony status, each T individual in group $i$ will receive cooperating benefit from its paired group $j$.
The inter-group interaction when $p_{ij}>0$ is shown in Fig.\ref{Fig.btinter}.

In Table 2, individuals' payoff of group $i$ are described as follows.
Individuals with A genotype(contributors) promote group's benefit by sacrificing their own interest.
Here, A individuals contribute interest $b$ to their group $i$ with cost $c$($b>c$). Each individual receives payoff $bf_i^A$ from the A individuals. Thus, PA individuals' payoff is $bf_i^A-c$, and PN with $bf_i^A$. TA and TN individuals' payoff is same as PA and PN respectively when a conflict occurs. However, if two groups are in a harmony status, TA and TN individuals receive another cooperating benefit $tmbf_j^T$ respectively from the T individuals of the paired group $j$. And $t$ is the benefit from one T individual of group $j$, $m$ for group capacity.

\begin{table}[H]
\centering
 \caption{Probability distribution table when group $p_{ij}>0$. column `win' means group $i$ or $j$ winning a conflict with a probability, `lose' for losing a conflict, `tie' for the conflict ending in a draw. $p_{ij}=f_i^{PA}-f_j^{PA}$ is the difference between $f_i^{PA}$ and $f_j^{PA}$.When $p_{ij}=0$, the conflict ends with a tie directly.}
\label{table0}
 \begin{tabular}{cccc}
  \toprule
  group & win & lose & tie\\
  \midrule
  $i$  & $0.5+0.5p_{ij}$ & $0.25-0.25p_{ij}$ & $0.25-0.25p_{ij}$  \\
  $j$  & $0.25-0.25p_{ij}$ & $0.5+0.5p_{ij}$ & $0.25-0.25p_{ij}$ \\
  \bottomrule
 \end{tabular}
\end{table}

\subsubsection{The intra-group interaction}
In the step of intra-group interaction, individuals will reproduce $m$ new offsprings. One individual can be selected as a parent according to this individual' share of the group's payoff, or say, individuals with greater payoff will have a greater chance to have more offsprings.
The new offsprings mutate in a probability $u$.

\begin{table}[H]
\centering
 \caption{Payoff table of group $i$. $n$ is the number of groups, $m$ for the individuals' number in each group, $d$ for the ratio of eliminated fighters(PA individuals) when a conflict occurs, $b$ for altruists contribute to public good at a cost $c$, $t$ for the benefit of harmonious relation, $u$ for the mutation probability.}
\label{table1}
 \begin{tabular}{ccc}
  \toprule
  $ $ & P & T \\
  \midrule
  A  & $bf_i^A-c$ & $bf_i^A-c+tmf_j^T$  \\
  N  & $bf_i^A$ & $bf_i^A+tmf_j^T$  \\
  \bottomrule
 \end{tabular}
\end{table}
\subsubsection{The update}
In the update part, the new offsprings replace their parental generation, then migrate to other groups randomly.
For the change of identity, it is not necessary to kill individuals if group members change their identities. Here, the new members who have changed their ethnic identity with probability $q$ migrates to a randomly chosen group.

\section{Simulation experiments and results}

In this section, we try to validate the proposed viewpoint: ethnic harmony could be promoted by popularizing civic identity. To validate this viewpoint, we should firstly test the proposition: that ethnic harmony(ethnic conflicts) has a positive(negative) association with civic identity.
Therefore, we will test whether the frequencies of ethnic conflicts has a negative correlation with the number of people with civic identity.
If the first proposition is wrong, the viewpoint could be wrong. However, if the proposition is right, secondly we will consider whether it is possible to keep a consistent peace by cleaning all P individuals or increasing a ratio of T individuals. If the second proposition is right, the solution for avoiding conflicts are given. However, if the second proposition is wrong, we will search a possible way to control the frequencies of conflicts in a low level.

Thus, the simulation experiments include three parts: 1)whether the number of people with civic identity has a positive relation with the level of ethnic harmony(i.e. has negative relation with the frequencies of conflicts); 2)whether it is possible to eliminate conflicts by cleaning P individuals completely, and whether it is possible to deduce ethnic conflicts by increasing T individuals; 3)how to promote between-groups' harmonious relationship, i.e., how to control the conflict frequency in a low level.
The related parameters in the simulations are listed in table \ref{table2}.

\begin{table}[H]
\centering
 \caption{Parameters of simulation experiments. $n$ is the number of groups, $m$ for the capacity of each group, $d$ for the ratio of eliminated fighters(PA individuals) when a conflict occurs, $e$ for the ratio of eliminated individuals in the losing group, $b$ for A individuals contributing benefit to their group and sacrificing their own benefit $c$, $t$ for the benefit from one T individual of the paired group when two paired groups are in harmonious status, $u$ for the new offsprings' mutation probability, $q$ for new offsprings' migrating probability.}
\label{table2}
 \begin{tabular}{ccccccccc}
  \toprule
  $n$ & $m$ & $d$ & $e$ & $b$ & $c$ & $t$ & $u$ & $q$\\
  \midrule
  18  & 30 & 0.2 & 2.5 & 0.05  & 0.03 & 0.001 & 0.005 & 0.2\\
  \bottomrule
 \end{tabular}
\end{table}

\subsection{The association between the ratio of  T individuals and conflicts}

In this section, the experiments focus on the association between T and the ratio of conflicts. The variation trend between the ratio of T individuals and conflicts in 1000 generations are shown in Fig.4. $f_{t}$ is the ratio of T individuals in one generation, $f_{c}$ for the ratio of occured conflicts. Moreover, the quantitative association between $f_{c}$ and $f_{t}$ of Fig.4 is listed in table 4.

\begin{figure}[H]
 \centering
  \subfigure[The relation between $f_{c}$ and $f_{t}$ when $f_{t}$ changes in generation 1-1000]{
    \label{Fig.3:subfig:a} 
    \includegraphics[width=7cm,height=4cm]{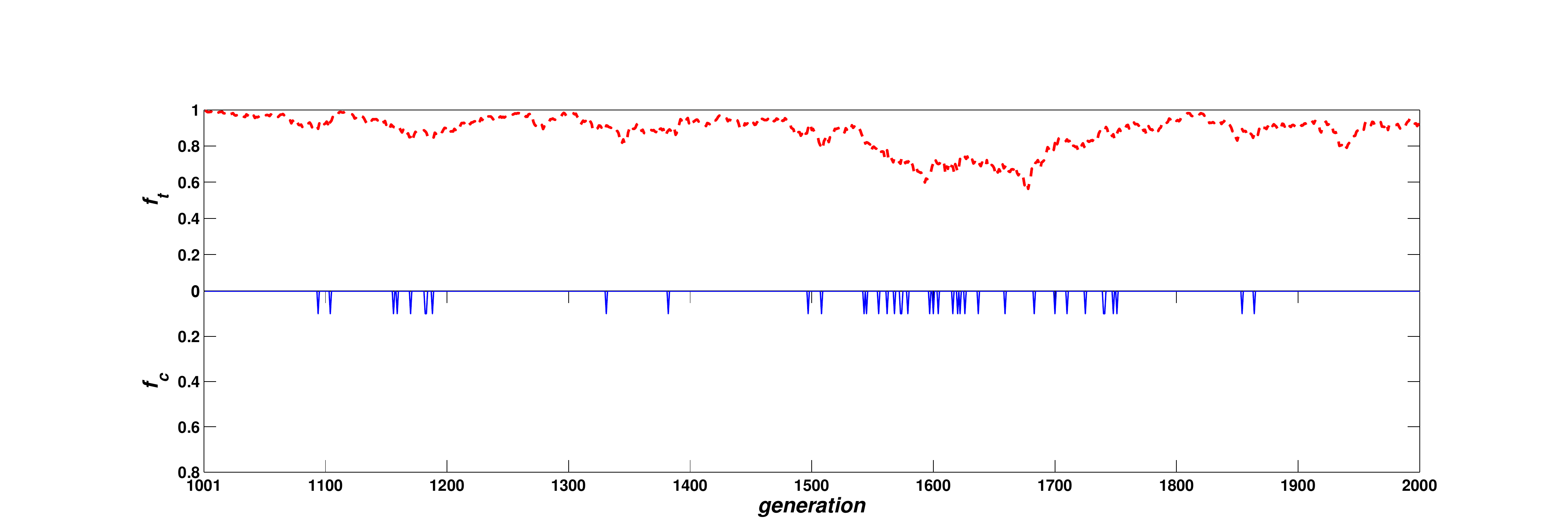}}
  \hspace{1cm}
  \subfigure[The relation between $f_{c}$ and $f_{t}$ when $f_{t}$ changes in generation 1001-2000]{
    \label{Fig.3:subfig:b} 
    \includegraphics[width=7cm,height=4cm]{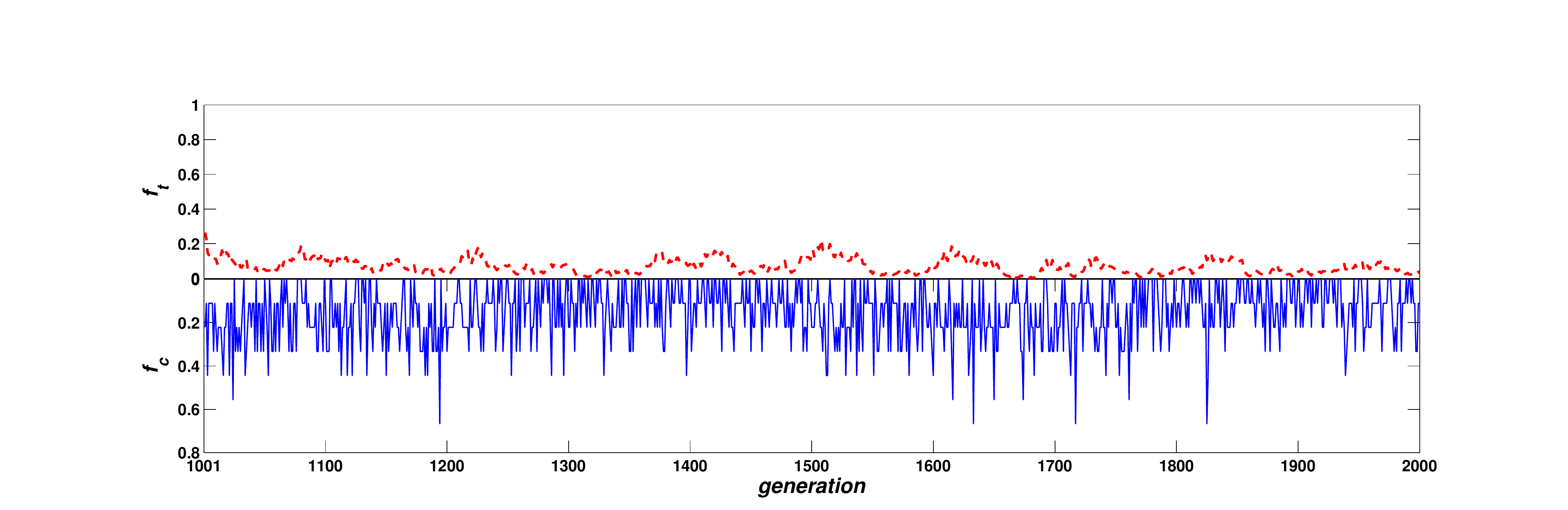}}
    \hspace{1cm}
  \caption{Association between $f_{t}$ and $f_{c}$. The horizontal axis $generation$ indicates the evolved $ith$ generation. The upper vertical axis $f_{t}$ is the ratio of T individuals of one generation, and the bottom vertical axis $f_{c}$ for the ratio of occured conflict frequencies account for maximum conflict frequencies in one generation. The red dot-dash line in the upper panel represents $f_{t}$, and The blue curve for $f_{c}$.}
\label{Fig.3} 
\end{figure}
Fig.4 shows that $f_{t}$ has a negative correlation with $f_{c}$. Because the level of $f_{c}$ is low when $f_{t}$ is great in Fig.4(a), and vice verse for Fig.4(b).
Correspondingly, the average value of $f_{t}$ and $f_{c}$ in 200 generations are denoted as $\overline{f_{t}}$ and $\overline{f_{c}}$ in Table 4.
The groups in Fig.4(a) are in a harmony status, because $\overline{f_{c}}$ is very small- around 0.008. Conversely, the groups in Fig.4(b) are in frequent conflicts because of $\overline{f_{c}}$ going up to 0.383 maximally.


\begin{table}[H]
\centering
 \caption{The quantitative association between $\overline{f_{c}}$ and $\overline{f_{t}}$ of Fig.4. Here, $Experiment groups$ are the simulated results in Fig.4, $generations$ for continued 200 generations in Fig.4, $\overline{f_{t}}$ and $\overline{f_{c}}$ for the average $f_{t}$ and $f_{c}$ in 200 generations respectively.}
\label{table3}
 \begin{tabular}{cccc}
  \toprule
  $Experiment groups$ & $generations$ & $\overline{f_{t}}$ & $\overline{f_{c}}$ \\
  \midrule
  Fig.4(a) & 1001-1200  & 0.937 & 0.008 \\
  Fig.4(a) & 1201-1400  & 0.923 & 0.002 \\
  Fig.4(a) & 1401-1600  & 0.856 & 0.012 \\
  Fig.4(a) & 1601-1800  & 0.782 & 0.016 \\
  Fig.4(a) & 1801-2000  & 0.909 & 0.002 \\

   Fig.4(b) & 1001-1200  & 0.084 & 0.383 \\
   Fig.4(b) & 1201-1400  & 0.065 & 0.252 \\
   Fig.4(b)& 1401-1600  & 0.081 &  0.297 \\
   Fig.4(b) & 1601-1800  & 0.059 & 0.364 \\
   Fig.4(b) & 1801-2000  & 0.056 & 0.229 \\
  \bottomrule
 \end{tabular}
\end{table}

\subsection{Effect on the ratio of conflicts by cleaning all P or increasing T}
From the results in section 3.1, we know that the ratio of T individuals has negative correlation with the ratio of occured conflicts. So whether is it possible to improve harmonious groups' relationship by eliminating all P individuals or increasing T individuals?
The following simulation experiments on cleaning all P individuals and increasing T individuals are shown in Fig.5.

Fig.5 and Table 5 show an interesting phenomenon: the $f_{c}$ will decrease in a short term, but it will bounce back quickly in a few generations.
Thus, in the long generations, it cannot keep the conflicts in a low level by eliminating all P individuals or increasing T individuals.

Specifically, by eliminating all P individuals in Fig.5(a) in generation 3550, conflicts decrease apparently in the following a few generations, yet it bounces back in generation 3650. The similar results happened in Fig.5(b) and Fig.5(c). The decrease of $f_{c}$ in Fig.5(b) is not as apparent as that of in Fig.5(c), because Fig.5(c) increases more T individuals than Fig.5(b).
The average ratio of conflicts before and after the operation(cleaning P or increasing T) in 100 and 200 generations are listed in table 5.
In Table \ref{table4}, $\overline{f_{c1}}$ is the average $f_{c}$ in 100 generations before the operation, $\overline{f_{c2}}$ for the average $f_{c}$ in 100 generations after the operation, $\overline{f_{c3}}$ for the average $f_{c}$ in 200 generations after the operation.
In table 5, we can know that $f_{c}$ has a decrease in the following 100 generations and bounce back again in the next 200 generations.
After cleaning all P individuals in generation 3550 in Fig.5(a), $\overline{f_{c2}}$ decreases to 0.2 obviously compared with the original average level $\overline{f_{c1}}=0.360$. Then $\overline{f_{c3}}$ increases to 0.35, which is almost close to the original level. Similarly, $\overline{f_{c2}}$ of Fig.5(b) and Fig.5(c) has a slight decrease, and the $\overline{f_{c3}}$ of Fig.5(b) bounces back to the original level nearly. Moreover, $\overline{f_{c3}}$ of Fig.5(c) bounces higher compared with the original level.

\begin{figure}[H]

  \subfigure[Clear all P individuals in generation 3550]{
    \label{Fig.5:subfig:a} 
    \includegraphics[width=8cm,height=4cm]{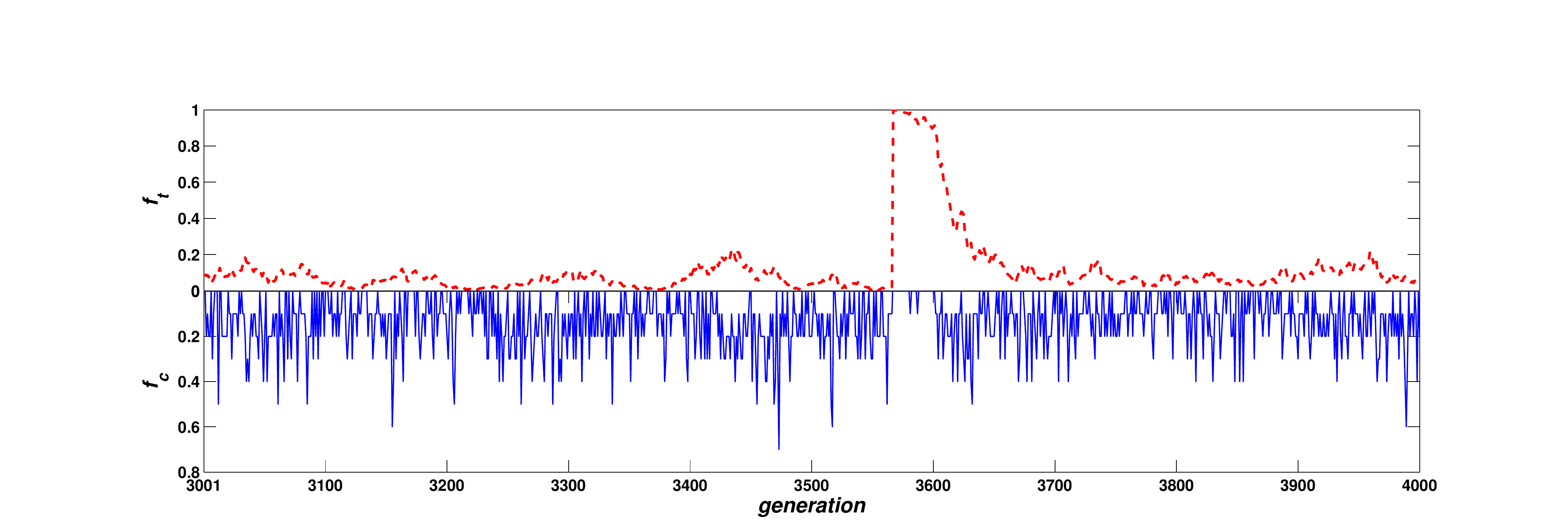}}
  \hspace{1cm}
  \subfigure[Increase 40\% T individuals in generation 3550]{
    \label{Fig.5:subfig:b} 
    \includegraphics[width=8cm,height=4cm]{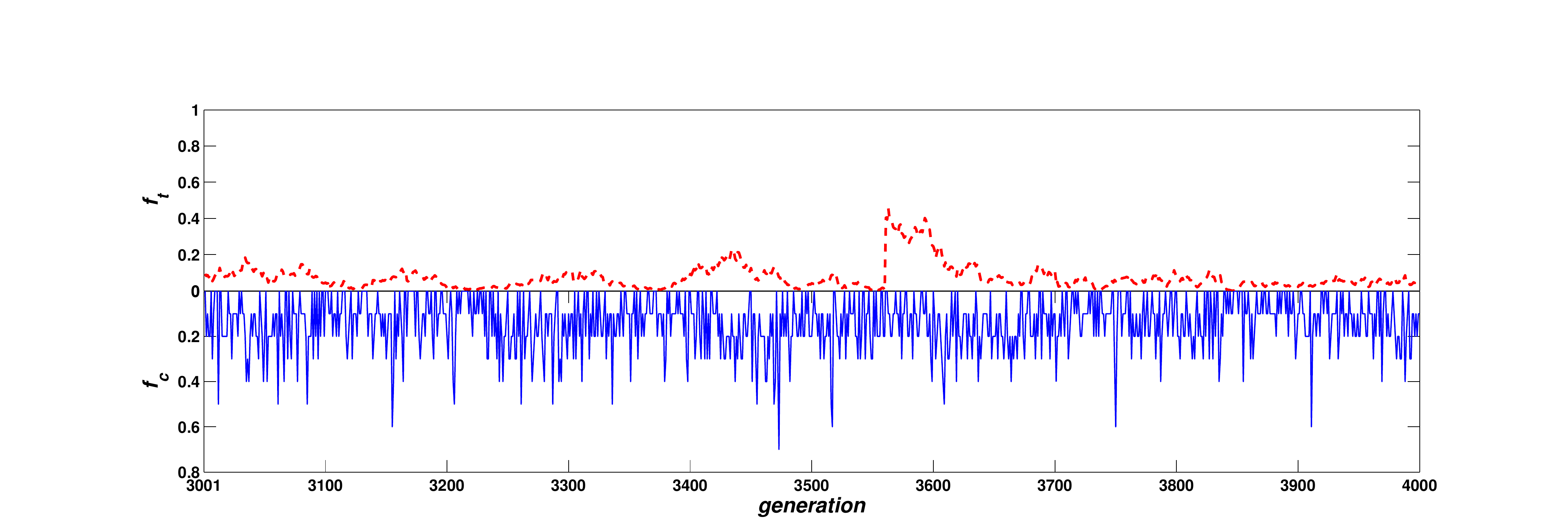}}
    \hspace{1cm}
  \subfigure[Increase 60\% T individuals in generation 3550]{
    \label{Fig.5:subfig:c} 
    \includegraphics[width=8cm,height=4cm]{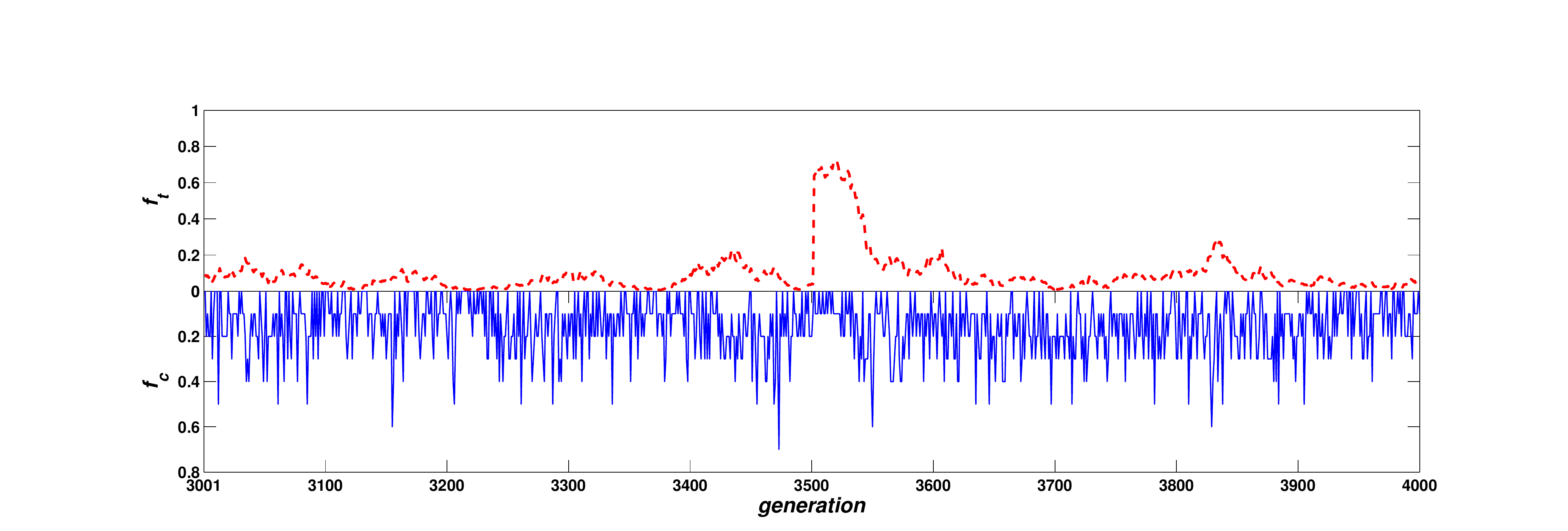}}
  \caption{The effect on $f_{c}$ by cleaning all P individuals, increasing 40\% T, 60\% T individuals respectively.}
  \label{Fig.4} 

\end{figure}

\begin{table}[H]
\centering
 \caption{ The quantified effect on $f_{c}$ with the increasement of T or eliminating P individuals.}
\label{table4}
 \begin{tabular}{ccccc}
  \toprule
  $Experiment groups$ & $\overline{f_{c1}}$ & $\overline{f_{c2}}$ & $\overline{f_{c3}}$ \\
  \midrule
 Fig.5(a)  & 0.360 & 0.200 & 0.350  \\
 Fig.5(b)  & 0.372 & 0.348 & 0.350  \\
 Fig.5(c)  & 0.385 & 0.336 & 0.398  \\
  \bottomrule
 \end{tabular}
\end{table}

\subsection{Increasing T periodically to reduce between-groups' conflicts}
Because of the rebound, the conflict ratio cannot be controlled by the force of increasing T individuals once for all.
What the result will be if we increase more T individuals before the rebounding point?
Experiments with increasing T individuals periodically are shown in Fig.6. Also, the corresponding $\overline{f_{t}}$(average $f_{t}$ in 200 generations) are listed in table 6.
\begin{figure}[H]

  \subfigure[Increase 40\% T individuals periodically each 200 generations]{
    \label{Fig.5:subfig:a} 
    \includegraphics[width=8cm,height=4cm]{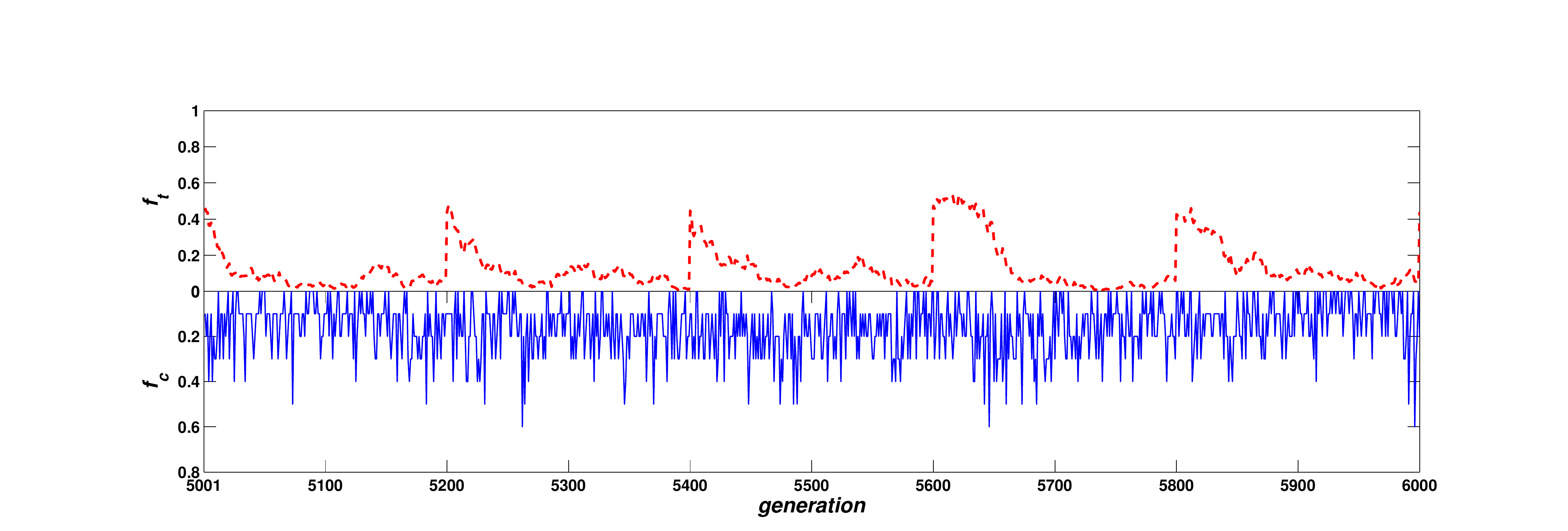}}
  \hspace{1cm}
  \subfigure[Increase 40\% T individuals periodically each 50 generations]{
    \label{Fig.5:subfig:b} 
    \includegraphics[width=8cm,height=4cm]{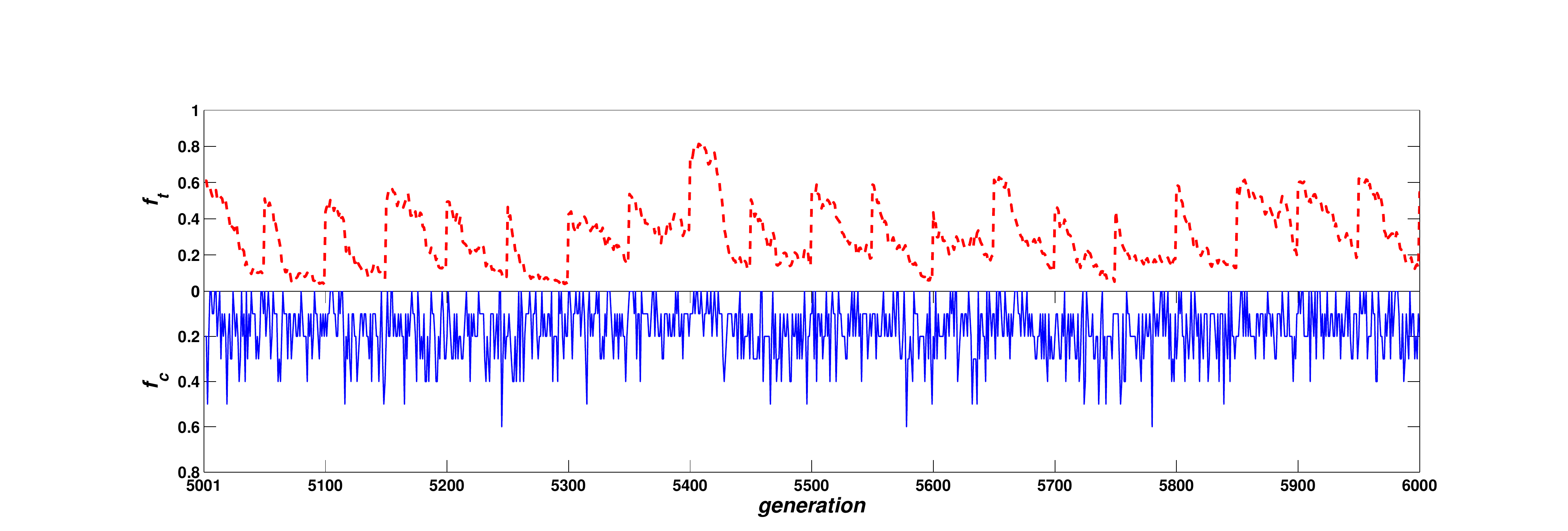}}
    \hspace{1cm}
    \subfigure[Increase 60\% T individuals periodically each 50 generations]{
    \label{Fig.5:subfig:b} 
    \includegraphics[width=8cm,height=4cm]{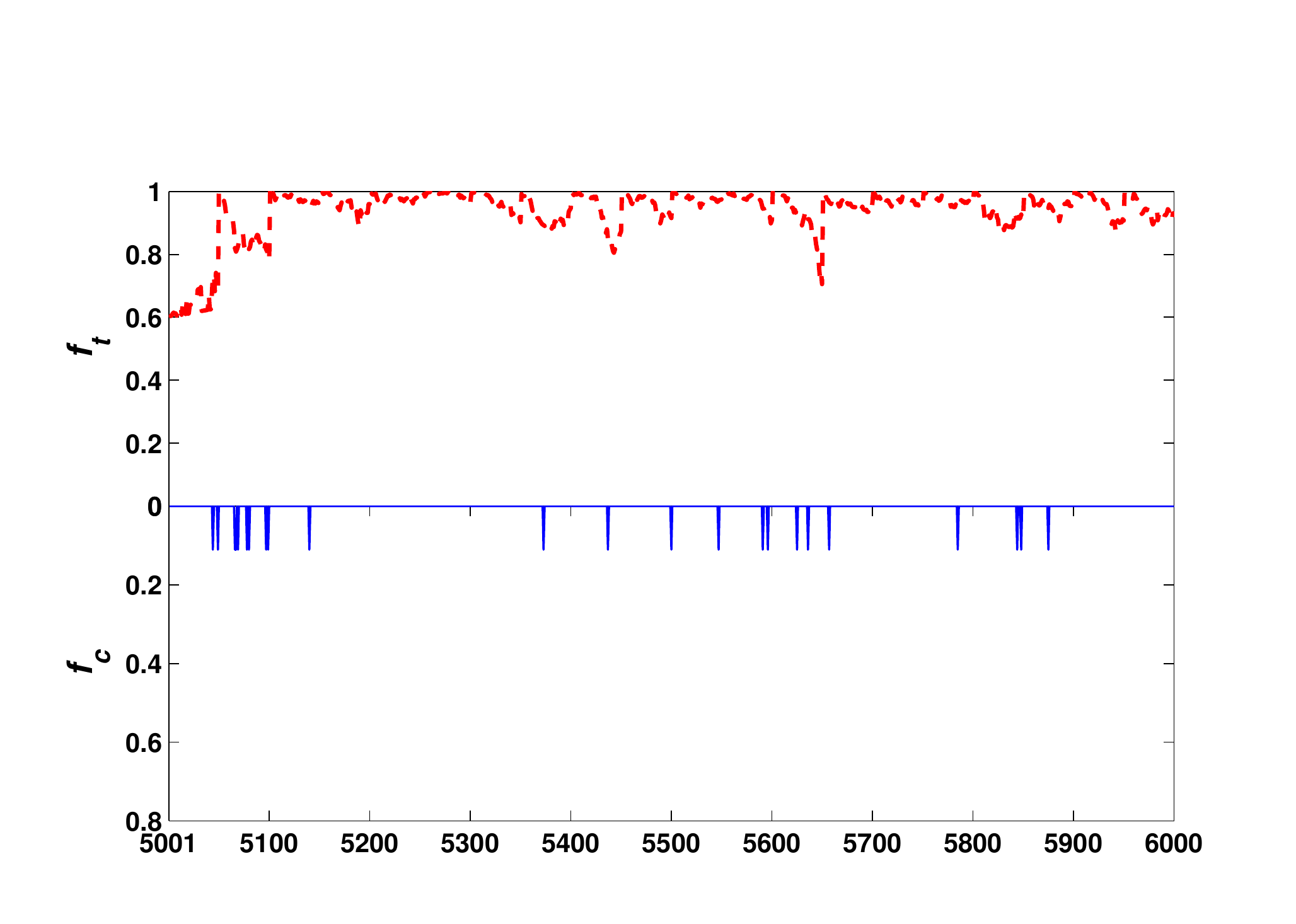}}
  \caption{The effect on $f_{c}$ by increasing 40\% T, 60\% T individuals periodically.}
  \label{Fig.5} 
\end{figure}

\begin{table}[H]
\centering
 \caption{The quantified effect on $f_{c}$ with the increase or decrease of $f_{t}$}
\label{table5}
 \begin{tabular}{cccc}
  \toprule
  $Experiment groups$ & $generations$ & $\overline{f_{t}}$ & $\overline{f_{c}}$ \\
  \midrule
  Fig.6(a) & 5001-5200  & 0.280 & 0.344 \\
  Fig.6(a) & 5201-5400  & 0.269 & 0.317 \\
  Fig.6(a) & 5401-5600  & 0.336 & 0.346 \\
  Fig.6(a) & 5601-5800  & 0.257 & 0.368 \\
  Fig.6(a) & 5801-6000  & 0.377 & 0.315 \\

  Fig.6(b) & 5001-5200  & 0.477 & 0.281 \\
  Fig.6(b) & 5201-5400  & 0.367 & 0.314 \\
  Fig.6(b) & 5401-5600  & 0.359 & 0.294 \\
  Fig.6(b) & 5601-5800  & 0.552 & 0.243 \\
  Fig.6(b) & 5801-6000  & 0.560 & 0.249 \\

  Fig.6(c) & 5001-5200  & 0.97 & 0.001 \\
  Fig.6(c) & 5201-5400  & 0.964 & 0.005 \\
  Fig.6(c) & 5401-5600  & 0.945 & 0.004 \\
  Fig.6(c) & 5601-5800  & 0.966 & 0.001 \\
  Fig.6(c) & 5801-6000  & 0.957 & 0.004 \\
  \bottomrule
 \end{tabular}
\end{table}

In Fig.6, we increase 40\% and 60\% T individuals every 200 and 50 generations respectively.
From Fig.6(a) and Fig.6(b), we can see that $f_{c}$ bounce back with the period, because the period is too long or the increased T individuals are not enough, or both.
But we can control $f_{c}$ in a low level by increasing T individuals before the bounce.  In Fig.6(c), the $f_{c}$ keeps in a very low level by increasing 60\% T individuals each 50 generations. By this way, the $\overline{f_{c}}$ of Fig.6(c) is no more than 0.05, and it is the smallest in the three figures.

So, under the same increasement of T individuals, the smaller the period the lower level of $f_{c}$ is. Moreover, under the same period, the lager the $f_{t}$ then the lower level of $f_{c}$ is.
\subsection{Sensitivity analysis}
In this section, we will analyze the effect on $f_{c}'$ and $f_{t}'$ under different parameters. $f_{c}'$ and $f_{t}'$ are the average value of $f_{c}$ and $f_{t}$ in 10000 generations.
In the simulations, we change one parameter and fix other parameters, and the simulations are listed in Fig.7.
From Fig.7, we can see that the value of $f_{c}'$(or $f_{t}'$) could be fluctuant when the value of one parameter changes. However, the responding $f_{c}'$ is small when $f_{t}'$ become great. So the negative association between the frequencies of conflicts and the number of T individuals still holds. Therefore, the other two conclusions in section 3.2 and 3.3 that base on the negative association also hold.

Specifically, parameters $m$(group's capacity), $c$(altruists' sacrificed benefit), $q$(migrating probability) and $t$(the benefit from the paired group's T individuals under harmonious status) affect $f_{c}'$ and $f_{t}'$ a lot. In Fig.7(a),(e),(f),(h), when $m$, $c$, $q$ and $t$ become large, the corresponding $f_{c}'$ tend to be small. The reasons are stated as follows: when $m$ is large, the probability of leading to a conflict becomes small because of the large population base; when $c$ becomes large, the $f_{c}'$ become small because the number of T individuals will increase(the PA individuals' fitness become small); when $q$ is large, the $f_{c}'$ become small because the conflicts occur with a low probability; when $t$ is large, individuals tend to be friendly with other groups, then the $f_{c}'$ will become small.
On the contrast, when $b$ becomes large, the $f_{c}'$ tend to become large. Because the large population of altruists could increase the number of fighters(PA individuals).
The change of other parameters $d$,$e$ and $u$ have no direct association with the increase or decrease of $f_{c}'$.

The simulations enlighten us that it also exists other possible ways to decrease the frequencies of conflicts. For examples, conflicts could be decreased by increasing the chances of communication between groups(increasing $q$), or by promoting the groups' cooperating benefit(increasing $t$) etc..
\begin{figure}[H]

  \subfigure[]{
    \label{Fig.m:subfig:a} 
    \includegraphics[width=8cm,height=4cm]{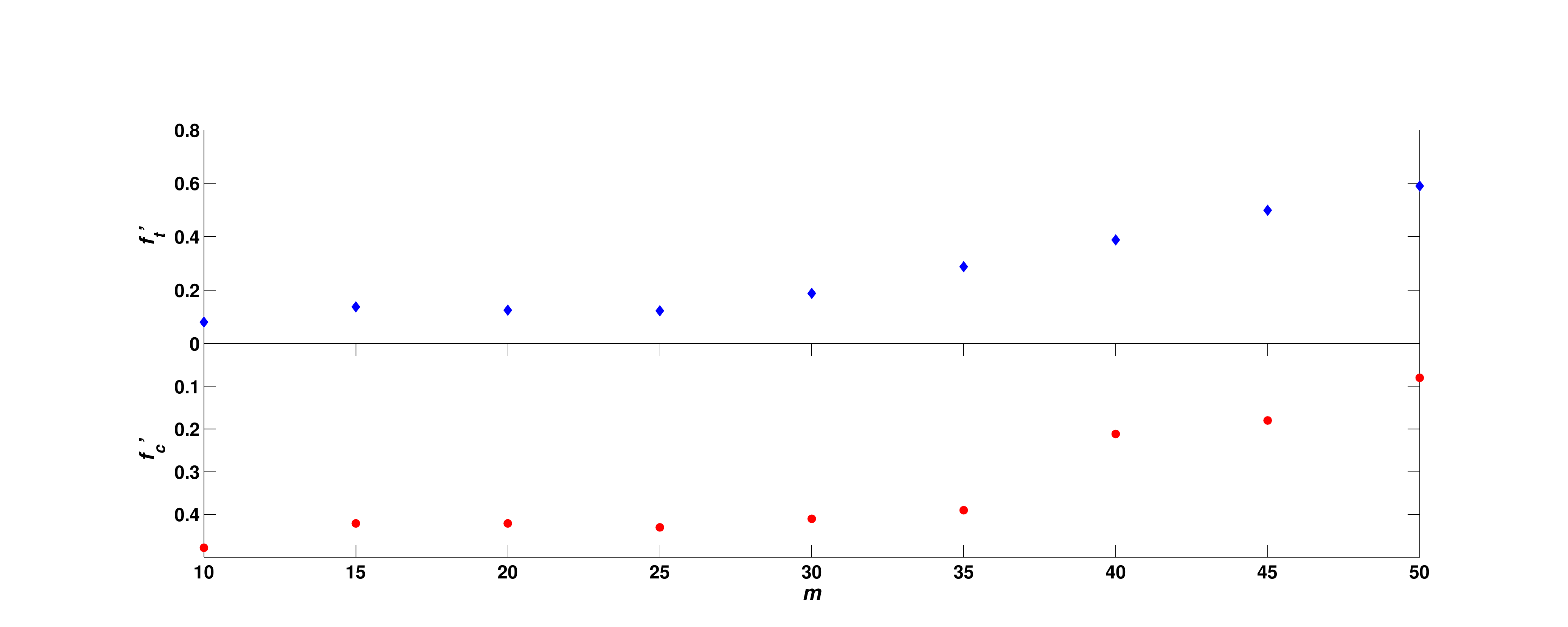}}
  \hspace{1cm}
  \subfigure[]{
    \label{Fig.d:subfig:b} 
    \includegraphics[width=8cm,height=4cm]{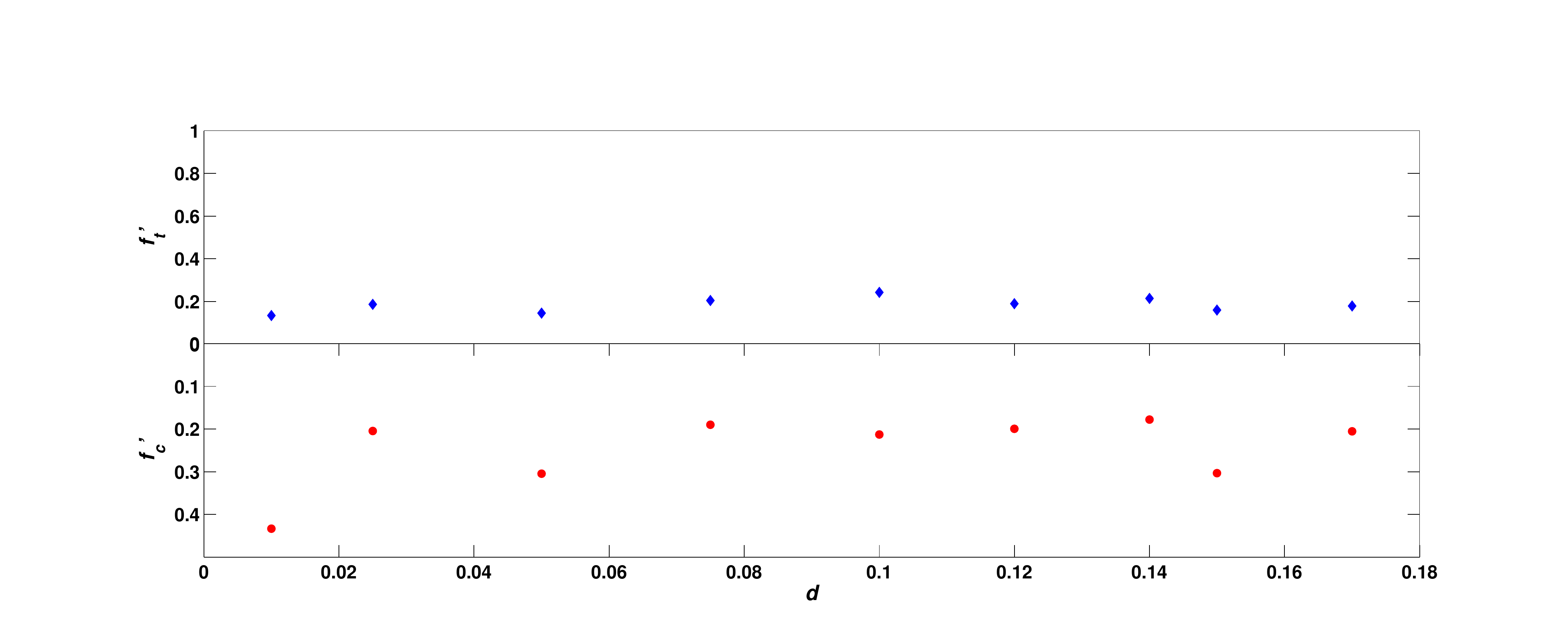}}
    \hspace{1cm}
    \subfigure[]{
    \label{Fig.e:subfig:b} 
    \includegraphics[width=8cm,height=4cm]{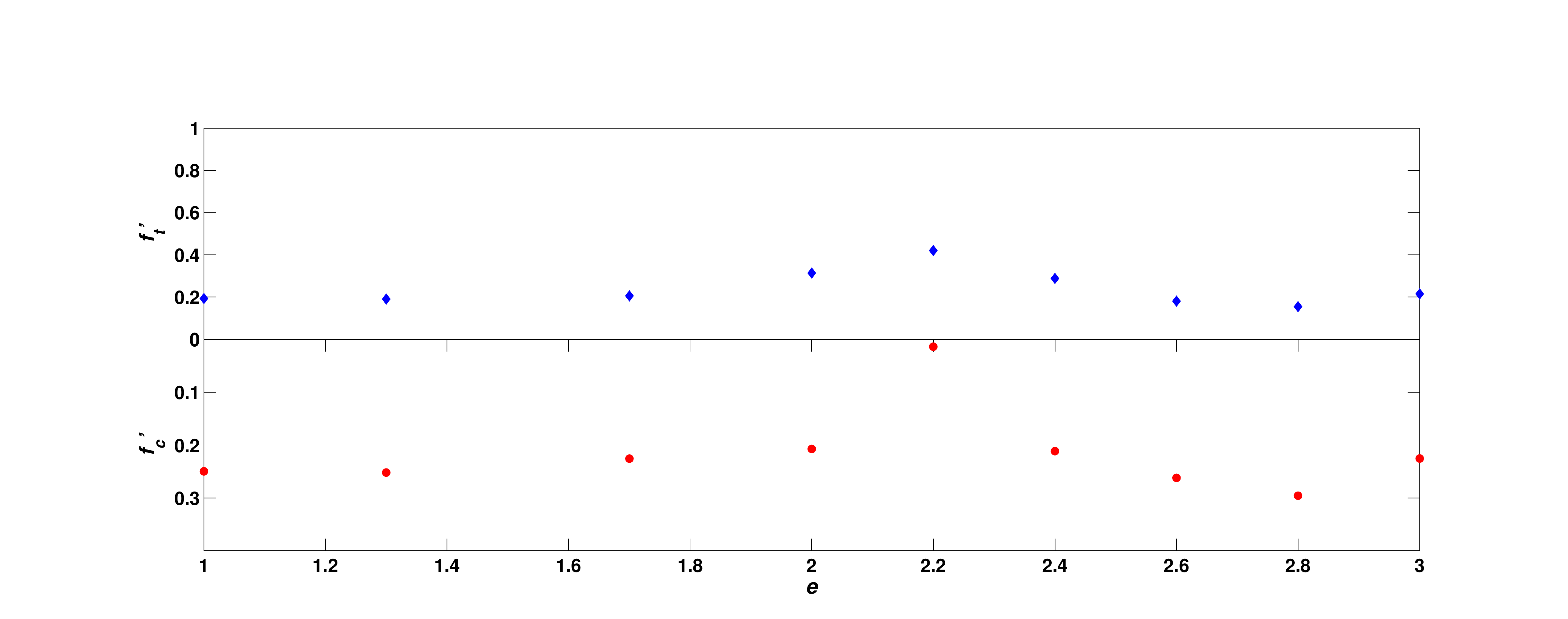}}
    \hspace{1cm}
      \subfigure[]{
    \label{Fig.b:subfig:a} 
    \includegraphics[width=8cm,height=4cm]{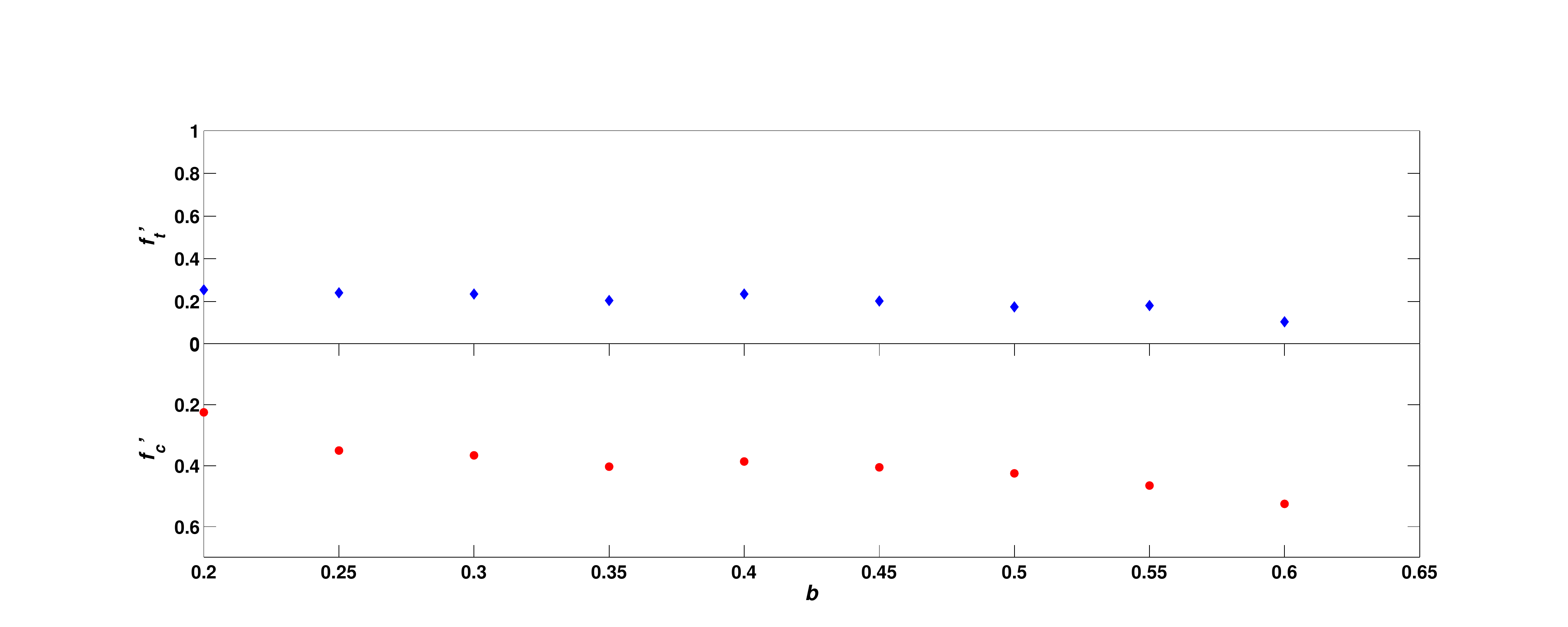}}
  \hspace{1cm}
    \subfigure[]{
    \label{Fig.c:subfig:a} 
    \includegraphics[width=8cm,height=4cm]{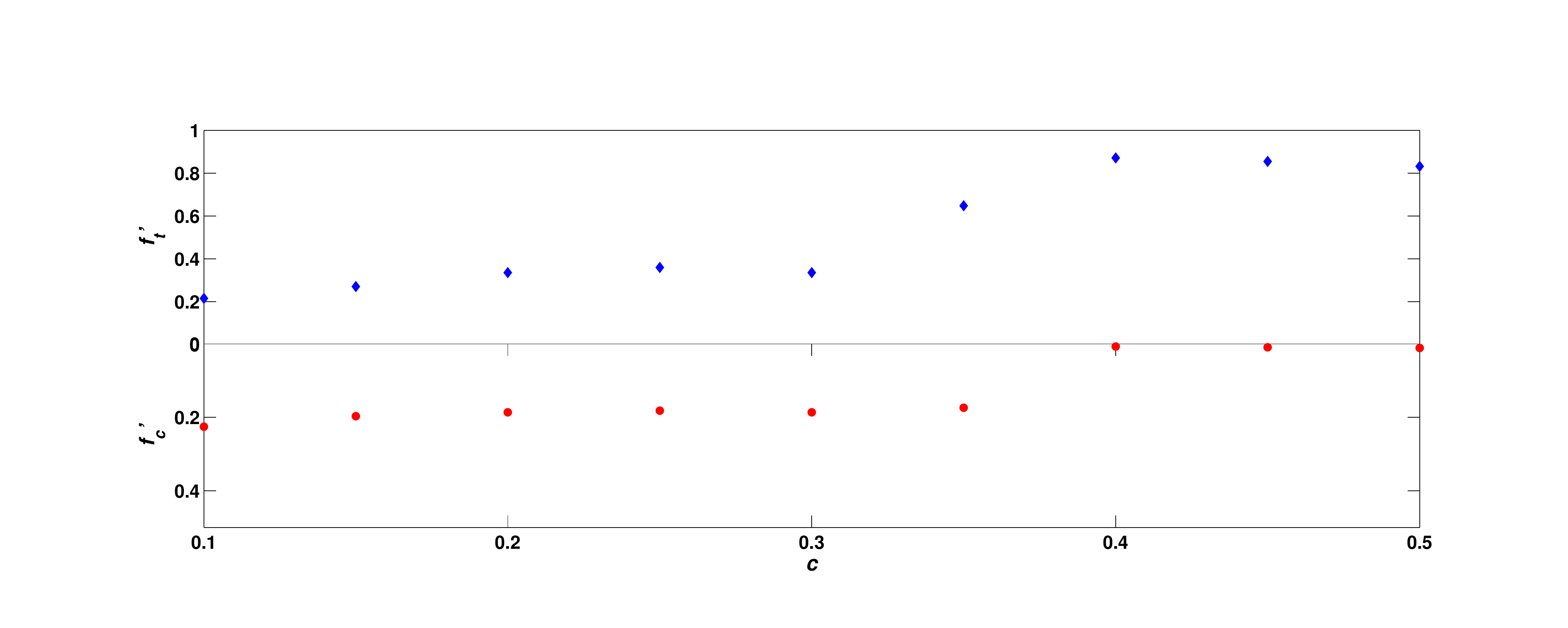}}
    \hspace{1cm}
    \subfigure[]{
    \label{Fig.t:subfig:a} 
    \includegraphics[width=8cm,height=4cm]{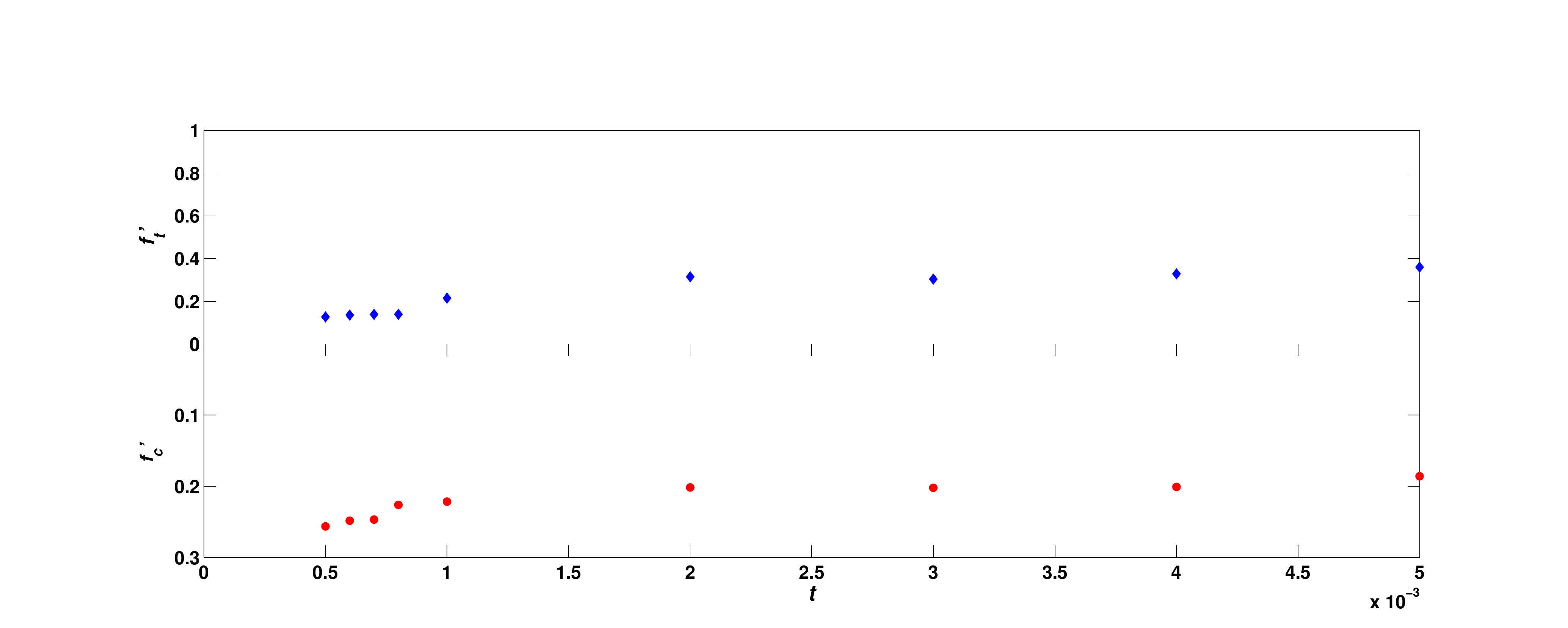}}
    \hspace{1cm}
    \subfigure[]{
    \label{Fig.u:subfig:a} 
    \includegraphics[width=8cm,height=4cm]{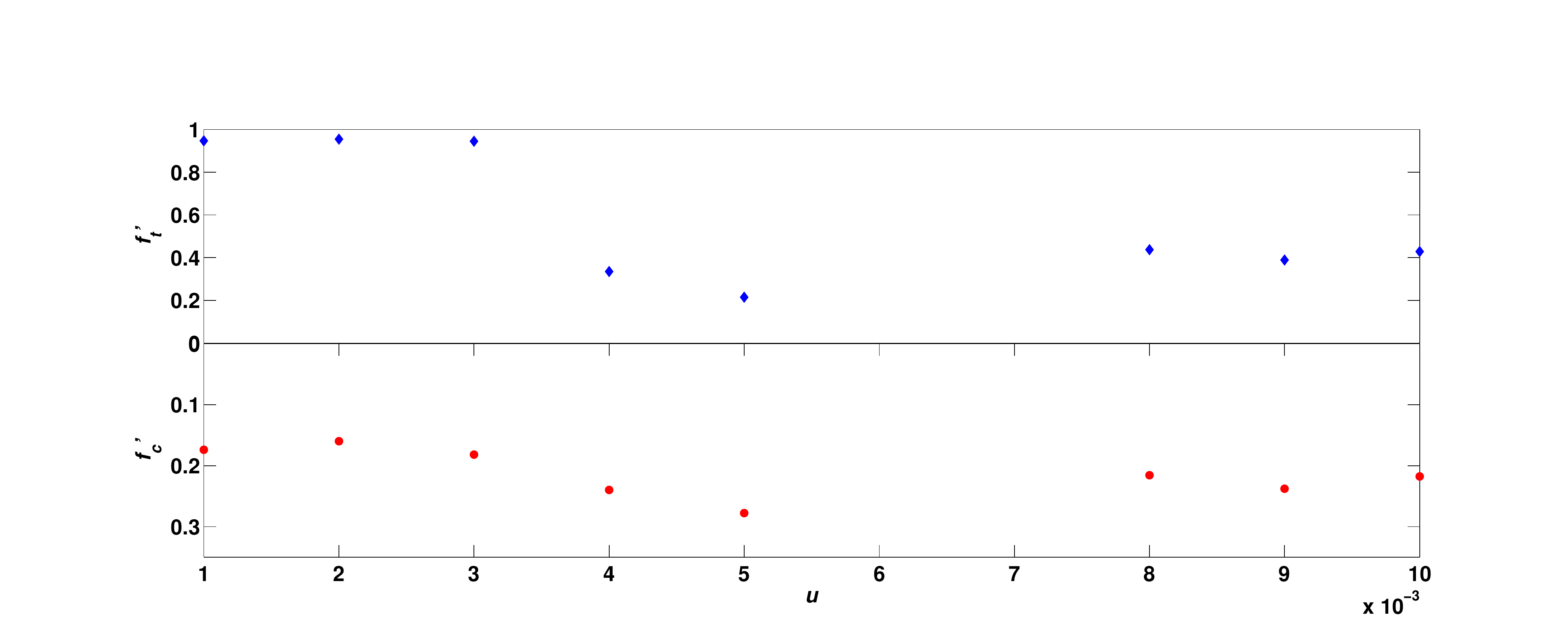}}
    \hspace{1cm}
    \subfigure[]{
    \label{Fig.q:subfig:a} 
    \includegraphics[width=8cm,height=4cm]{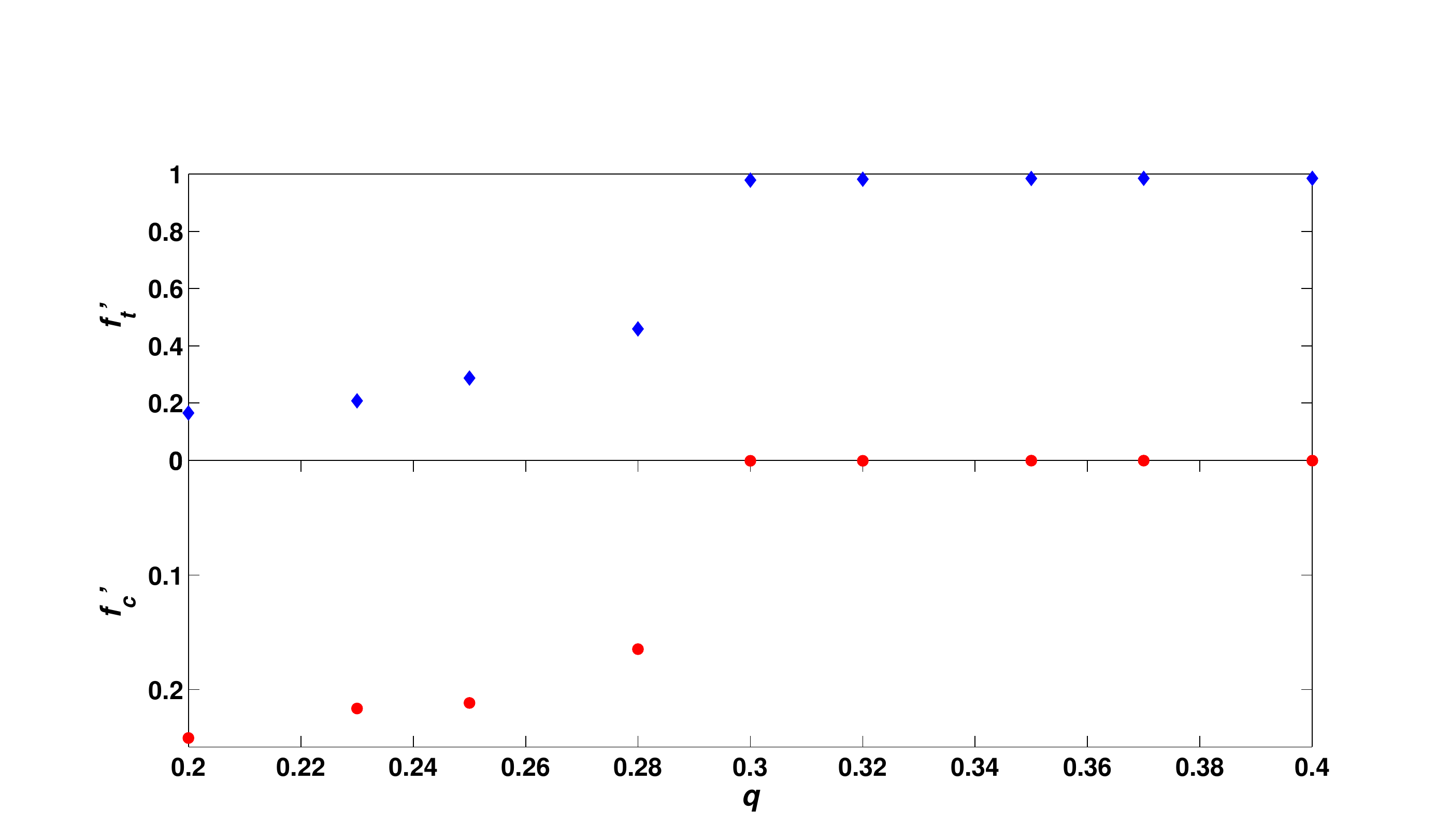}}

  \caption{Sensitive analysis of Parameters.}
  \label{Fig.5} 
\end{figure}

\section{Conclusion}
In this paper, we established an evolutionary game model, and simulated the relationship between civic identity and ethnic conflicts by the evolutionary game theory.
The simulation results indicated that ethnic conflicts cannot be avoided by just killing the ultranationalist. Though the ratio of individuals with civic identity had a negative association with the ratio of ethnic conflicts, ethnic conflicts cannot be kept in a low level in a long term just by promoting civic identity, because the ration of conflicts may bounce back. However, the frequency of conflicts could stay in a low level by propelling civic identity persistently and periodically.

Indicated by the simulation results, we have known that it is possible to control ethnic conflicts in a low level by promoting civic identity. However, to predict ethnic conflicts is still a challenge, because it is hard to collect the historical datasets about ethnic identity and ethnic conflicts. The effects of popularizing civic identity on ethnic groups' culture, language are still needed to explore.
Moreover, this work spawns some interesting problems. Can the evolution of national languages be studied by the methodology of natural science- evolutionary theory and stimulation technology? What the interesting findings can be found by studying the evolutionary of language with evolutionary theory and stimulation technology?
What's more, the study in this paper could supply an example for studying ethnic problems by the methodology of natural science.

\section{Acknowledgments}
The authors are grateful for support from the Fundamental Research Funds for the Central Universities(No.CZY12032), Natural Science Foundation of Guangxi(No.2011GXNSFB018074), Scientific research project of the Guangxi Education Department(No.200911lx406,200103YB136) and the State Key Laboratory of Networking and Switching Technology(No.SKLNST-2010-1-04).
BZ is grateful for support from the State Key Laboratory of Software Engineering (No.2012-09-15).





\end{document}